\documentclass{aa} 

\usepackage{graphicx,url,natbib}
\usepackage{txfonts}
\usepackage{hyperref}
\usepackage{mathrsfs}
\usepackage{color} 

\bibpunct{(}{)}{;}{a}{}{,}    
\makeatletter
\newcommand{\bibnote}[2]{\global\@namedef{#1note}{#2}}
\newcommand{\biblink}[2]{\global\@namedef{#1link}{#2}}

\newcommand{\unit}[1]{\ifmmode \:{\rm #1}\else \mbox{#1}\fi}

\newcommand{\MSun}{\,\mathrm{M_{\odot}}}
\newcommand{\nbody}{\textit{N}-body }

\newcommand{\eg}{\textit{e.g. }}

\newcommand{\dcut}{$d_{\textrm{cut}}\ $}
\newcommand{\calS}{$\cal S\ $}
\newcommand{\calN}{$\cal N\ $}

\makeatother

\usepackage[normalem]{ulem}

\begin{document} 

   \title{Assessing membership projection errors in star forming regions}   
   \author{T. Roland\inst{1}, 
           C.M. Boily\inst{1}, 
           L. Cambr\'esy\inst{1}}
   
   \institute{Universit\'e de Strasbourg, CNRS,
                  Observatoire Astronomique de Strasbourg, UMR 7550,
                          F-67000 Strasbourg, France\\
              \email{timothe.roland@astro.unistra.fr}\\
             }

   \date{Received 6 August 2020 / Accepted 19 October 2020}

 
  \abstract
   {Young stellar clusters harbour complex spatial structures emerging from the star formation process. Identifying stellar over-densities is a key step in better constraining how these structures are formed. The high accuracy of distances derived from \textit{Gaia} DR2 parallaxes still do not allow us to locate individual stars within clusters of  $\approx 1\, \rm{pc}$ in size with certainty.}
   {In this work, we explore how such uncertainty on distance estimates can lead to the misidentification of membership of sub-clusters selected by the minimum spanning tree (MST) algorithm. Our goal is to assess how this impacts their estimated properties.}
   {Using \nbody simulations, we build gravity-driven fragmentation (GDF) models that  self-consistently reproduce the early stellar configurations of a star forming region. Stellar groups are then identified both in two and three dimensions by the MST algorithm, representing respectively an inaccurate and an ideal identification. We compare the properties derived for these resulting groups in order to assess the systematic bias introduced by projection and incompleteness.}
   {We show that in such fragmented configurations, the dynamical mass of groups identified in projection is systematically underestimated compared to those of groups identified in 3D. This systematic error is statistically of $50\%$ for more than half of the groups and reaches $100\%$ in  a quarter of them. Adding incompleteness further increases this bias.
   }
   {These results challenge our ability to accurately  identify sub-clusters in most nearby star forming regions where distance estimate uncertainties are comparable to the size of the region. New clump-finding methods need to tackle this issue in order to better  define the dynamical state of these substructures.
   }

\maketitle
%

\section{Introduction}
Young stars are predominantly observed in the dense parts of spiral arms of disc galaxies. After a decade of \textit{Herschel} observations, it is now established that these star forming regions exhibit highly sub-structured configurations due to fragmentation of the gas in extended filaments \citep{andre2010, pokhrel2018}. 
The latest interferometric facilities (\eg ALMA, NOEMA) allow to study filaments and primordial cores at sub-parsec scales (see \eg \citealt{hacar2017, hacar2018, andre2019, montillaud2019}), bringing to light the very last stage of the transformation of gas into stars. 
Stars emerge from this network in groups of varying size and morphology. However, their kinematics is at odds with that of their natal gas \citep{foster2015}. Moreover, their dynamical state is poorly constrained \citep{kuhn2019} and their fate uncertain.

The new era of precise astrometry opened by \textit{Gaia} DR2 enables exploration of the spatial and kinematic properties of a huge number of young stellar systems (see \eg \citealt{kounkel2019, cantat-gaudin2019} for the newest catalogs of Gaia-discovered stellar clusters). In the solar neighbourhood, the 3D structure of local star forming clouds is within our reach. \cite{grossshedl2018} measured the extended 3D structures of the Orion A tail, reporting an estimated total length of $90\, \mathrm{pc}$, in significant disagreement with the projected size of $40\, \mathrm{pc}$. Other large, extended structures have been identified outside the solar circle \citep[\eg][]{alves2020}. The astonishing accuracy of \textit{Gaia} DR2 data peaks at $M_G \approx 13$, but degrades steadily at fainter magnitudes. This leads to uncertainties in distance estimates affecting mostly low-luminosity or reddened stars. 
In young and embedded regions, identifying 3D stellar structures at (sub-)parsec scales remains challenging.
For instance, the selected sample of \cite{kounkel2018} in Orion has a mean uncertainty on parallax measurements of $\approx0.5\, \mathrm{mas}$ which corresponds to $\approx8\, \mathrm{pc}$ at this distance ($\approx 400\, \mathrm{pc}$). 
        
In this context, simulations are key to testing our understanding of the processes acting in these fragmented regions. 
Recent computations of hydro-gravitational fragmentation and accretion, coupled with radiation-transfer physics, give promising results in terms of  for example the shape of the stellar mass function (\eg \citealt{bate2012}), formation of proto-planetary discs (\eg \citealt{hennebelle2016}), and the role of stellar feedback (\eg \citealt{wall2020}). Despite continued progress, the number of young stars produced in these high-resolution calculations is limited to a few hundred. Larger calculations remain challenging. Moreover, strong interactions between newly born stars occur in these simulations. 
Exploring the interplay between several of these systems is key to probing the formation of more massive clusters. Consequently, pure \nbody modelling is still of great use to explore configurations with $N>1000$ stars. This approach works well when gravity takes over magneto-thermal support and drives the fragmentation of the giant molecular cloud (GMC). This allows us to address specific topics such as primordial mass segregation \citep{pavlik2019} or the impact of gradual star formation \citep{farias2019}. As massive clusters are also likely to form massive stars \citep{motte2018}, \nbody simulations are also suited to exploring the environment where they form. 

There is a long-standing history of modelling stellar interactions with gravity alone. One major challenge of such work consists in generating realistic initial configurations. Different methods have been developed. One of the most straightforward is to follow snapshots of hydro-dynamical simulations and the structures traced by sink particles (\eg \citealt{moeckel2010, fujii2016}). This approach is well suited to exploring the long-term evolution of the original hydro-dynamical simulations, but is limited by the high computational effort to model the full star formation process. An efficient alternative that can be used to reproduce the fragmented aspect of star forming complexes  at lower cost was introduced by \cite{goodwin2004}, namely the box fractal method. This technique can recreate spatial configurations of various observed clusters and allows the user to tune the different levels of substructures. However, the velocity field 
is not constrained by the history of formation of such structures which makes the dynamical state of the system somewhat ad hoc. 

Another method providing a balance between computational cost and realistic configurations was proposed by \cite{dorval2016}, which we refer to as gravity-driven fragmentation (GDF) models. This method is an efficient way to build  fragmented initial conditions self-consistently using only \nbody modelling. It allows the user to explore a large number of fragmented configurations and test our understanding of the dynamical processes in action \citep[see also][]{dorval2017}. 

In this paper, we specifically discuss morphological biases introduced when estimating properties of stellar over-densities in a fragmented cluster seen in projection. We used the GDF method to model a realistic distribution of stars in a young star cluster (\S\ref{model}). Our methodology consists in identifying stellar groups of more than $50$ members (also referred to as clumps), both in 3D and in projections (2D) and estimating their properties (\S\ref{detection}). We used the virial dynamical mass to compare groups identified by both methods and assess the deviation introduced by projection effect (\S \ref{results}). Finally, we discuss the implications of our results in \S\ref{discussion} and conclude in \S\ref{conclusion}.


\section{Building realistic stellar distribution} \label{model}
\subsection{Gravity-driven fragmentation modelling} \label{GDF}

We start our study using the GDF models described in \cite{dorval2016}. We summarise here the motivations and principles of such models.

The fragmentation of a continuous cold fluid bound by gravity will proceed from continued growth of a spectrum of perturbations. When dealing with stars, the point-mass nature and limited number of elements means that the potential is not smooth, and Poisson-seeded density fluctuations rapidly develop into dense substructures. The growth of small-scale modes will proceed on very short timescales, however large-scale modes will grow on the global gravitational timescale of $\sim 1/\sqrt{G\rho_0}$ which is the same as the free-fall timescale. To allow time for these fluctuations to develop well in the non-linear regime, one sets up a self-gravitating system uniformly populated by stars with an outward radial velocity field. The velocity flow is taken to be isotropic with respect to the barycentre of the system; its amplitude is such that the total binding energy remains negative, and so expansion stops after some time which we take to coincide with the formation timescale of massive stars. 

\begin{figure*}
\centering
\includegraphics[width=0.99\textwidth]{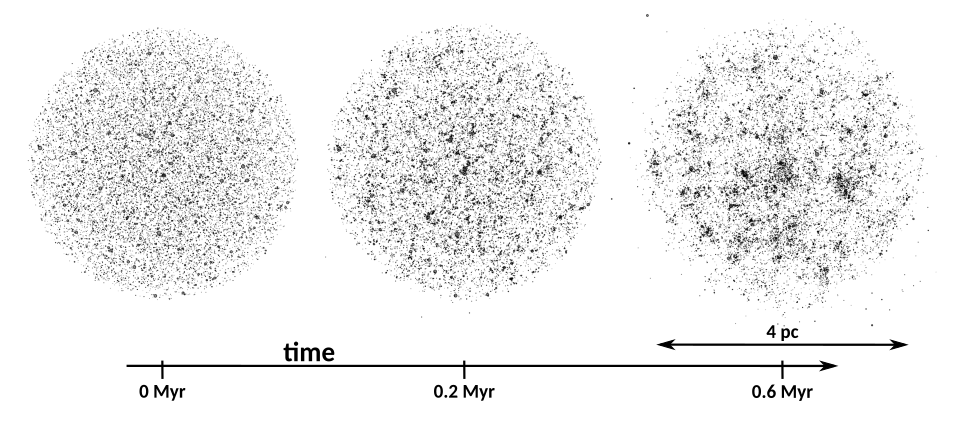}
\caption{Gradual fragmentation of a GDF model in co-moving coordinates (physical length scale is only valid for the right panel). The left panel shows the initial step: stars are uniformly distributed and have radial velocities. The middle panel shows the growth of fluctuations in the expanding sphere. The right panel shows the point at which the system is fully fragmented and expansion stops.}
\label{H-L}
\end{figure*}

This method allows us to simulate the two main physical processes driving the late stages of star formation: self-gravity, computed by a collisional \nbody integrator; and adiabatic cooling of the system, driven by the expansion. We adopted such a modelling to reproduce the distribution of young stars in a filamentary network whose nodes are dense and asymmetric clumps. These clumps comprise between tens of stars and a few hundred, and are usually considered as the early products of the star formation process \citep{krumholz2014}.
As the overall velocity field is self-consistently built, we are able to probe their dynamical state and interactions with their surroundings.
Furthermore, when using a mass spectrum, the most massive stars attract the surrounding low-mass stars to themselves. \cite{dorval2016} showed that massive stars are the seeds for the growth of fragment modes. As a result, clumps are mass segregated at birth, retrieving a key feature of hydro-simulations \citep[\eg][]{maschberger2011, wall2019}. They also harbour a top-heavy mass profile whose Salpeter index is comparable to the one found in young- and open clusters of the Milky Way \citep{bastian2010}.
Figure \ref{H-L} shows the initial, intermediate, and final steps for one of our models. More details and further examples are given in \cite{dorval2016} and \cite{dorval2017}.

\subsection{Simulation details}

We based our study on a GDF model with $N=15\,000$ young stars following a canonical Kroupa-Chabrier initial mass function \citep[IMF; ][]{kroupa2001, chabrier2003}. We adopt the functional form of \cite{maschberger2013} for efficiency. The initial velocity field is chosen so that the expansion stops roughly when the system is $4$~pc in diameter, resulting in a mean density of \[ \rho_0 \approx 150\ \rm M_\sun/pc^. \]
These parameters probe the high-mass regime of young clusters close to the so-called young massive clusters (YMCs). They were chosen for several reasons.
\begin{itemize}
    \item Firstly, these parameters allow us to accurately sample the IMF in the range $0.01\rm{-}30\, \rm M_\sun$. The high-mass end limit was chosen in order to keep the mean mass of the system around $\approx 0.3\MSun$. Indeed a specific draw of the IMF on $15\,000$ stars leads to approximately ten stars more massive than $30\MSun$\footnote{The probability to draw a star more massive than $30\,\rm M_\sun$ is of the order of $10^{-4}$ with $N \sim 10^4$.} and can contribute as much as $10\%$ of the mass budget of the overall region.

    \item Secondly, these allow us to reproduce a well-defined sample of clumps of different memberships and densities. This is a direct consequence of including enough massive stars that act as the seeds of the formation of these clumps. In a theoretical `bottom up' scenario of star formation, where these clumps will subsequently merge together, such a system is a likely progenitor of a massive open cluster. This large range in densities (see \S\ref{HOP}) allows us to probe different clump environments. However, we note that such a crowded configuration can affect the level of contamination. In \S\ref{lowN}, we discuss a model with lower membership.

    \item Finally, with such parameters, the formation of the model lasts $0.55\, \rm{Myr}$ which matches the typical time for massive stars ($>1\MSun$) to reach the main sequence. This also corresponds to the free-fall time of a GMC with density equal to $\rho_0$,
    \begin{equation}
    t_{\rm ff} = \sqrt{\frac{3\pi} {32G\rho_0}}=0.67\, \mathrm{Myr}.     \end{equation}
    As most of the stars in a GMC are formed within a free-fall time \citep{grudic2018}, this allows us to study the very early stages of the configuration of newly born stars. These structures have not yet been erased by two-body relaxation, and their dynamical state is a direct result of the fragmentation process.

\end{itemize}

We perform the computations with the \texttt{AMUSE} platform \citep{portegies2018}. We used the \nbody code \texttt{ph4} which is a fourth-order Hermite scheme integrator coupled with the \texttt{Multiple} module to handle efficiently close interactions \citep{hut1995, mcmillan1996}.
The relative energy drift $\delta E / E$ at each time-step is always lower than $10^{-6}$ except when close encounters cannot be resolved by the \texttt{Multiple} module, and can lead to an error up to $10^{-4}$. However, these remain rare, which makes the total energy drift during the whole calculation of the order of $10^{-4}$. We do not explicitly include any binary population but they are able to form dynamically through the fragmentation process. However, the fraction of multiple stars remains below $5\%$ and these do not influence our conclusions.

\subsection{Sub-selection within the GDF model} \label{HOP}

Gravity-driven fragmentation models allow exploration of the structures built by the fragmentation process. In this study, we specifically target clumps that can be misidentified when only seen in projection. However, as we see in Fig. \ref{H-L}, the spherical imprints of the initial conditions remain visible at the large scale. This overall geometry of the cluster can have an impact on the clumps that we identify in projection in the following section. 

To this end, we implemented a method to extract a subsample from the central parts of the system. This method is based on the \texttt{HOP} algorithm \citep{eisenstein1998} which was initially designed to identify over-dense haloes in any particle-based sample. 
\texttt{HOP} assigns to each particle a local density estimate based on a list of nearest neighbours and groups them around density peaks.
We adapted the \texttt{HOP} algorithm to select a sample representative of the complex structure of the GDF model spanning a large-enough range in density. First, we tune the parameters to select only particles with a local density above a given threshold. 
Several tests lead us to set the threshold value equal to  $\approx 75\, \rm M_{\sun}/pc^3$, or half of the median system density. We then exclude groups that were close to the edge of the GDF sphere by retaining only those that included at least one member lying less than $1$~pc from the centre of the model. This two-step procedure leads to the selection of a subsample of particles excluding about half of the stars.

\begin{figure}
        \centering
        \includegraphics[width=\columnwidth]{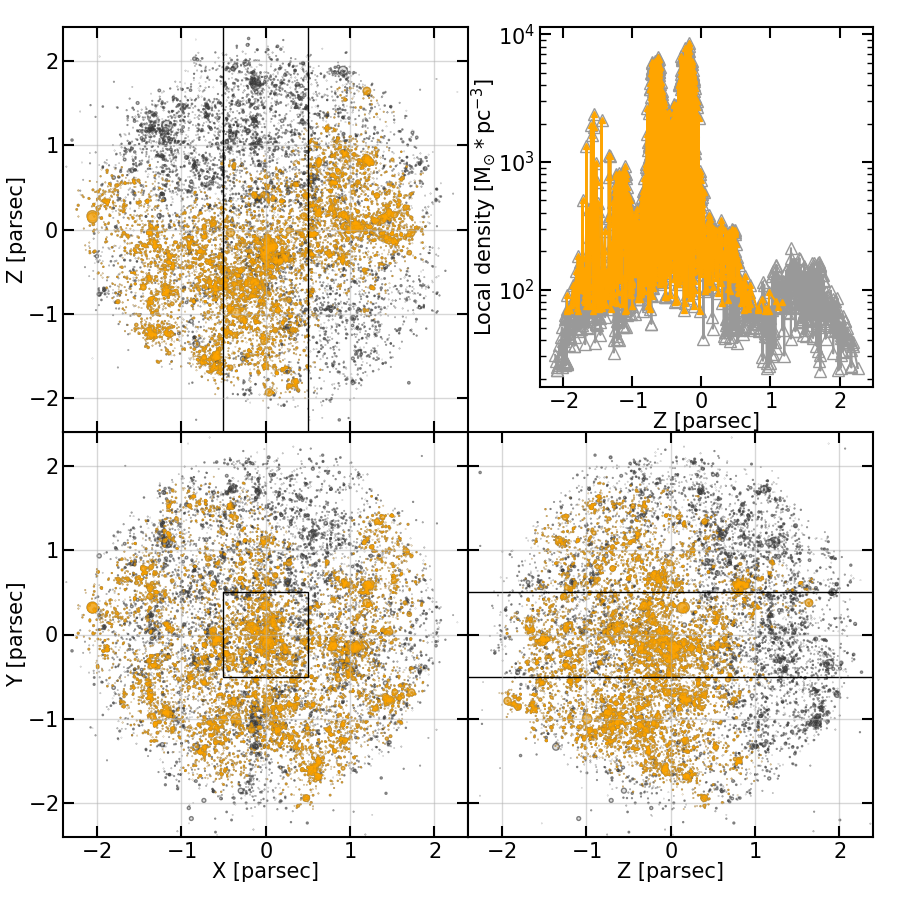}
    \caption{Three orthogonal views of the GDF model. The full initial sample of stars is represented in grey shades and the sub-sample we extracted is shown in orange. The upper right panel shows the local density of the particles estimated by \texttt{HOP} stacked along the slice represented by the black rectangles on the three other views.}
    \label{sub}
\end{figure}

This new sample is the reference set for the rest of the study. Figure \ref{sub} shows three orthogonal views of the initial GDF model (grey dots) and the selection of $7235$ stars (orange dots). On the upper left panel, local densities of particles within a slice of the model (black rectangles) are plotted. This highlights that the selection retains all the main density peaks. Furthermore, local densities span a large range, from $\approx100\, \rm \MSun/pc^3$ to $\approx10^4\, \rm \MSun/pc^3$. 

\section{Stellar groups} \label{detection}
Stars are more likely to be formed in dense environments where the final collapse of gas is transformed into stars through an accretion process. The various stages of accretion\footnote{These are known as Class 0, I, II, III objects \citep[see \eg][]{andre2000}} means that stars acquire their mass in a shared environment.
Data from \textit{Herschel} for example have long indicated that the proximity of young stars gives rise to extended structures of varying morphology, all interconnected by a complex network of bridges and filaments. We are tasked with identifying the stars that belong to such a network.

\subsection{Identification method}

Several techniques have been developed to detect over-densities in observations of star forming regions. These include for example, the minimum spanning tree (MST) algorithm \citep{gutermuth2009};  the finite-mixture model technique \citep{kuhn2014}; DBSCAN \citep{joncour2018}, and HDBSCAN \citep{kounkel2018}. Despite their specific features, most of these techniques rely on star number density estimates projected on the sky\footnote{There are a few exceptions, for instance: HDBSCAN cited above or the \textit{friends-of-friends} algorithm \citep{huchra1982} which both use spatial- and kinematic coordinates, but they are difficult to set up.}. This is mainly because of the few observables available in such crowded regions, where extinction may be significant. By construction, these methods may miss the true underlying 3D structure of the objects under study. 

We address the biases that arise from projection effects, focusing on the MST method. 
Its ability to detect over-dense structures of any shape makes it a powerful tool in  studies of star forming regions.
It has been widely used  by both theorists (\eg \citealt{maschberger2010}) and observers (\eg \citealt{kirk2011, beuret2017}). MST statistics can also be used to derive other properties of stellar clusters such as morphological aspects \citep{cartwright2004} or a measure of mass segregation \citep{allison2009}.
Moreover, the MST method may be adapted to treat either 2D or 3D space distributions. Our setup described below was similarly implemented in both cases. This uniformity in the methodology is crucial in order to avoid introducing additional biases between 2D and 3D identifications. 

\begin{figure}
        \center
        \includegraphics[width=.95\columnwidth]{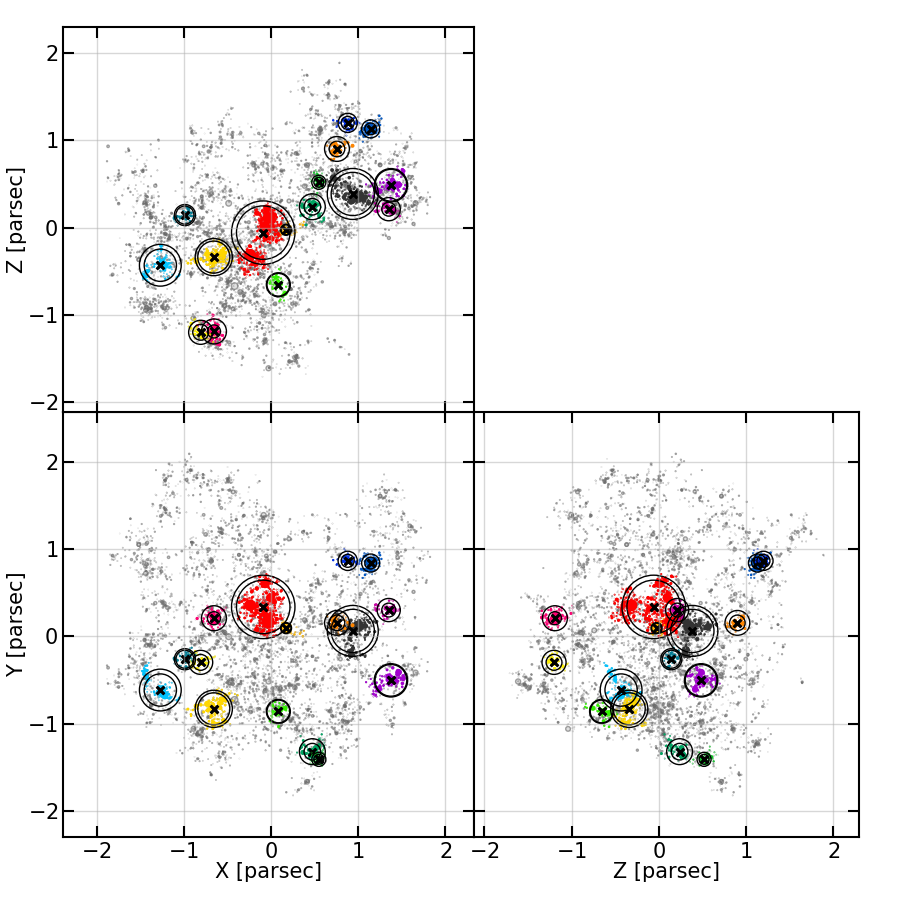}
        \caption{Three orthogonal views of the sub-sample with real groups identified by the MST in three dimensions. The black crosses show the location of the centre of mass of each group, and the circles show their $75\%$ and $90\%$ Lagrangian radii.}
        \label{MST3D}
        \includegraphics[width=.95\columnwidth]{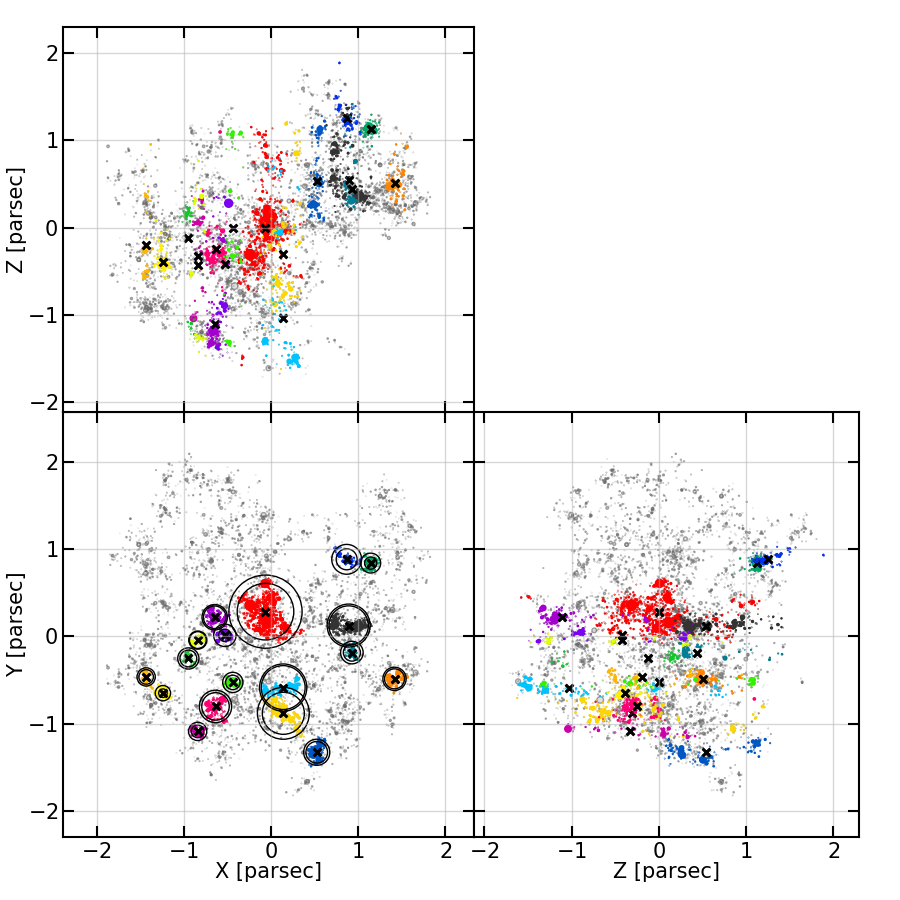}
        \caption{Same as Fig. \ref{MST3D} but with projected groups identified by the MST in two dimensions along the \textit{xy} plane (lower left panel). The projected Lagrangian radii are not tractable along the z-axis.}
    \label{MST2D}
\end{figure}

\subsection{Minimum spanning tree setup}

We summarise the MST procedure briefly (see \eg \citealt{gutermuth2009} for further details).
A spanning tree is a path linking all particles of a data set without forming closed loops. The MST is the spanning tree that minimises the total length of this path. We thus obtain a hierarchical set of edges that link every particle of the sample. To identify groups, we define a characteristic cut-off scale, $d_{\rm cut}$, such that all edges of larger values will be removed from the MST set. This isolates star aggregates that are on average closer to each other.
The choice of $d_{\rm cut}$ is critical to our analysis: a large value will link together stars that are very widely separated, revealing a few populous and sub-structured groups of stars ; on the other hand, a very small value will isolate binary and multiple stars, but will not reveal the hierarchical complex structure of the system as a whole. A natural way to pick a value for \dcut is to count the number \calS of subsets with a minimum membership, and vary $d_{\rm cut}$. When decreasing \dcut from $\approx R$, the size of the system, it is not too hard to see that \calS should rise from a value of $1$ to a maximum value $\mathcal{S}_{\rm{max}}$ and drop to $\approx 0$ when \dcut $\rightarrow 0$. 
An optimal value of \dcut is found when maximising the number of groups, \calS$=\mathcal{S}_{\rm{max}}$ \citep[see][]{dorval2016}. However, this iterative procedure can prove computationally costly. After some experimentation, we settled on a simple but efficient procedure based on the distribution of the MST edges. A characteristic length is retrieved by adding one-half of a standard deviation to the mean value of MST edges. The cutting length \dcut is set to 
\begin{equation}
  d_{\mathrm{cut}}=\mathrm{mean}(E_{\mathrm{MST}})+ \mathrm{std}(E_{\mathrm{MST}})/2,
\end{equation} 
with $E_{\mathrm{MST}}$ being the set of MST edges.
This way of fixing \dcut provides nearly optimal values without the iteration procedure. 

\subsection{Membership \cal N}
We fixed a minimum number of stars per stellar group, $\mathcal{N}$, based on two arguments. The first is that we consider a system that is on the order of $1\, \mathrm{Myr}$ old, which is old enough for massive stars to have reached the main sequence, but not so old that the system as a whole 
has had time to settle into virial equilibrium. Hence, filaments and knots are transient features. On the other hand, we want the majority of these clumps to be as close as possible to their state at formation: indeed, MST statistics cannot be used to gauge the equilibrium state of the substructures we aim to isolate. However, what is clear is that any substructure will evolve internally by phase mixing and kinetic-energy diffusion on a timescale of $\propto 1 / \sqrt{G\rho}$. We take as reference an average density of $100\gamma \MSun/\mathrm{pc}^3$, with $\gamma$ being a dimensionless scalar. We then compute a characteristic time $t_{\mathrm{cr}} \sim 1.5/\sqrt{\gamma}\times 10^6\, \mathrm{yr}$, which we take as the dynamical timescale for motion driven by gravity. When a number \calN of stars are bound together, their repeated interactions lead to diffusion of kinetic energy on a timescale $t_{\mathrm{rel}}$ given by \citep[\eg][]{meylan1997}:

\begin{equation}
    t_{\mathrm{rel}} \simeq 0.138 \left( \frac{R_{\rm{h}}}{2R_{\rm{v}}} \right)^{\frac{3}{2}} \frac{\cal N}{\ln 0.4\cal N} \, t_{\mathrm{cr}} \, .\end{equation}

In the above relation, the ratio of half-mass radius $R_{\rm{h}}$ to virial radius $R_{\rm{v}} = - \mathrm{G} M / E $ (where $E$ is the binding energy) is $\approx 1$ for self-gravitating configurations near equilibrium. 
We find $t_{\mathrm{rel}} > t_{\mathrm{cr}}$ for \calN $\gtrsim 50$, and furthermore $t_{\mathrm{rel}} > 1\, \mathrm{Myr}$ for $\gamma \le 1$. This tells us that clumps with membership \calN of 50 or more should remain relatively close to their phase-space configuration at birth. Even so, the case for setting \calN = 50 becomes weaker in denser environments $(\gamma \gg 1)$, and we should increase \calN in proportion to $\sqrt{\gamma}$ to avoid diffusion effects which drive dissolution. Clearly, a setup that will fit all situations is difficult to pin down; our reference value $= 50$ also coincides with the membership of stellar associations seen in the low-density environments \citep{lada2003, gouliermis2018}. This value is  also, in a loose sense, motivated by observations.

\begin{figure*}
        \centering
        \includegraphics[width=0.99\textwidth]{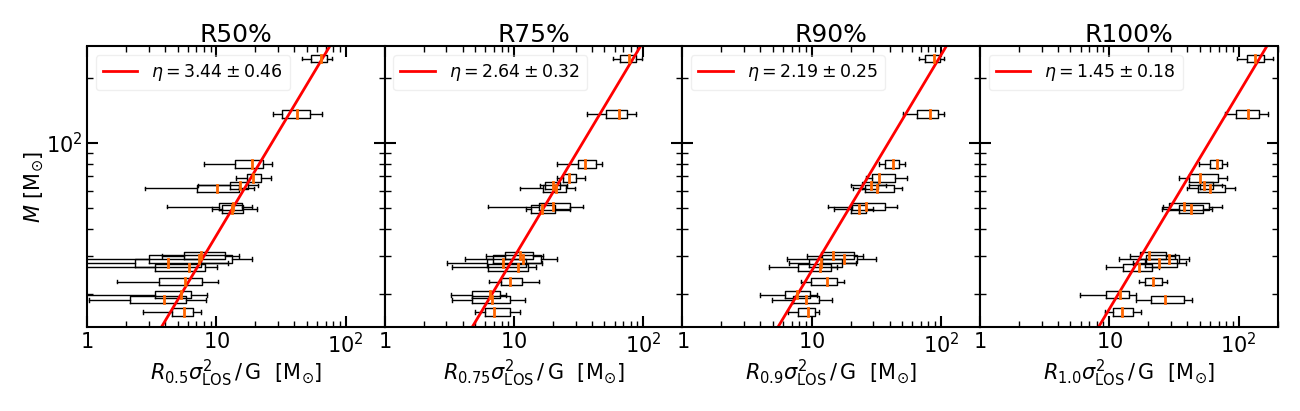}
    \caption{Correlation between the true and dynamical mass of real groups. Each panel shows different chosen radii for the dynamical mass estimate. All possible values along different projecting angles for the same group are shown in a box-plot: the orange vertical tick marks represents the median of the distribution; the edge of the box, the $16^{th}$ and $84^{th}$ percentiles (such that $68\%$ of the data are inside the box), and the horizontal bars correspond to the total width of the full distribution. The red lines are the least square fits on the data giving the values of $\eta$ corresponding to each radius.}
    \label{corr3d}
\end{figure*}

\begin{figure*}
        \centering
        \includegraphics[width=0.99\textwidth]{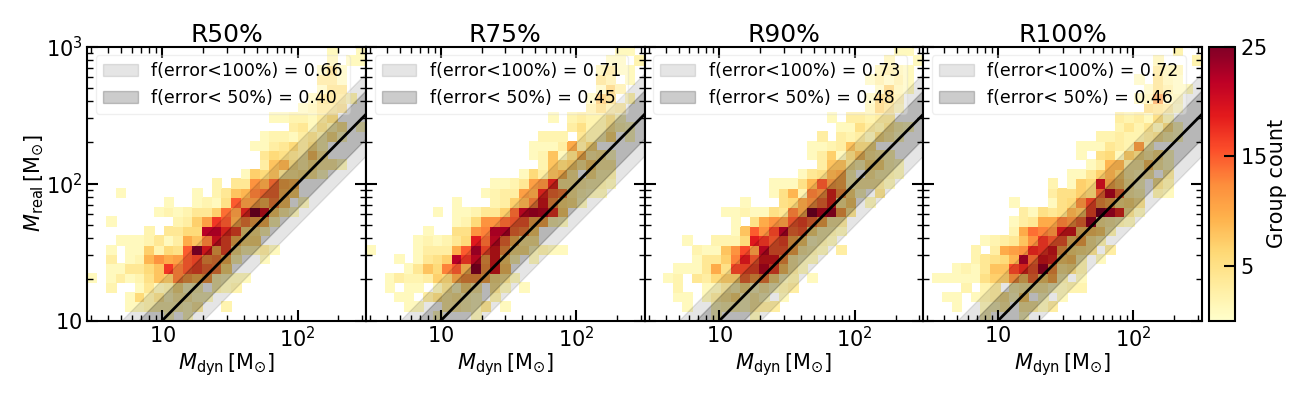}
    \caption{Correlation between the true and dynamical masses of groups identified in projection represented as a 2D histogram. The black line shows the expectation constrained from real 3D groups (see Fig. \ref{corr3d}). Different shaded areas show the fraction of well-estimated group masses within an error of $50\%$ and $100\%$}
    \label{corr2d}
\end{figure*}

\subsection{Dynamical mass estimates} \label{Mdyn}

We apply the MST algorithm to the subsample defined in \S\ref{HOP} twice: once using the full (3D) spatial coordinates, and once using only projected coordinates on the \textit{xy} plane. Three orthogonal views of the resulting groups identified in 3D are plotted in Fig. \ref{MST3D}, and in 2D in Fig. \ref{MST2D}. This leads to the identification of approximately $16$ groups in each case. However, the comparison between these figures shows a different identification. 
In Fig. \ref{MST3D}, the main over-densities of the region are well retrieved in all projection panels; we refer to these as \textit{real groups}, as they represent the ideal identification with the MST algorithm.
By contrast, in Fig. \ref{MST2D}, group members appear well clumped only in the \textit{xy} plane. On the other planes, they spread out over the whole area. These groups, henceforth \textit{projected groups}, are biased in various ways: they can be \textit{real} over-densities with some unrelated contaminants; or, completely artificial structures whose members are just aligned by chance in the line-of-sight. We emphasis the fact that this is different from the classical contamination by field stars, because all stars in our model are lying within the same star forming region of $\approx4\, \mathrm{parsecs}$ in diameter.
Real and projected groups cannot be compared one to one. In order to compare both sets statistically, we quantify an observable property of these groups self-consistently. 

With this in mind, we compute the dynamical mass which is a powerful tool to gauge the equilibrium state of a system \citep{spz2010}. It is also used to evaluate the dynamical state of young stellar systems and determine their fate \citep[see \eg][]{kuhn2019}.
From the virial theorem, we define the dynamical mass $M_{\rm{dyn}}$:
        \begin{equation}                        
                M_{\rm{dyn}} = \eta \times \frac{R_{\mathrm{eff}}  \sigma_{\rm LOS}^2}{G},
        \label{mdynproj}
        \end{equation}
where $R_{\rm eff}$ is the effective radius, and $\sigma_{\rm LOS}$ is the line-of-sight velocity dispersion. The mass-to-light ratio, denoted $\eta$, depends on the phase-space configuration of the stellar system \citep{boily2005, fleck2006}. Hence all quantities that appear on the right-hand side of Eq.~(\ref{mdynproj}) are observables from projected data.  
This is crucial to compare $M_{\rm{dyn}}$ for real and projected groups in a self-consistent way. 

As groups in our sample are asymmetric and not fully relaxed, we first determine the value of $\eta$ empirically on real groups. This will constitute a value of reference to assess the difference between real and projected groups (we discuss the value of $\eta$   further in \S\ref{eta}).

To obtain an ensemble of $\eta$s representative of each group, we estimate the dynamical mass of real groups from a large number of different lines of sight, representing the possible viewing angles all over the sphere. To pick isotropic lines of sight, we chose the \textit{HEALPix} tessellation \citep{gorski2005} with the resolution parameter $N_{\rm side} = 3$, giving $192$ different angles of view separated by $\approx14$ degrees each.
Our procedure iterates over these lines of sight and re-computes the position and velocity of the whole sample to estimate the projected dynamical mass of real groups corresponding to each view. Finally, since the \textit{HEALPix} tessellation is symmetric with respect to the centre of the sphere, each configuration is paired with its mirror image. As a 
result there are $192/2 = 96$ independent mass estimates for each real group.

The effective radius in Eq.~(\ref{mdynproj}) is often replaced with the half-light radius in observational surveys of star clusters. The half-light radius matches the half-mass radius of analytic King-Michie cluster models in equilibrium \citep{mengel2002, wolf2010} which allows direct comparison between observations and theory. 
However, this no longer holds true when the stellar clusters are mass-segregated \citep{gaburov2008}. 
Moreover, the MST identifies asymmetric groups for which only one radius may not be representative of its size. Therefore, we explored different definitions of group radius corresponding to the projected Lagrangian radii at $50\%$, $75\%$, $90\%,$ and $100\%$.

Computing the line-of-sight velocity dispersion of groups identified by the MST is not trivial. Indeed groups are in close interaction with their environment and members nearer to the edge are likely to be affected by the surroundings. 
We address this issue by estimating four different velocity dispersions corresponding to the stars lying within the four Lagrangian radii computed above.
We also chose to compute the velocity dispersion using the biweight scale estimator \citep{beers1990} which is more robust than the standard deviation when the distribution contains outliers. 

\section{Assessing projection bias} \label{results}

\subsection{Real groups}

We thus obtain four different sets containing $16\times96=1536$ different estimates of the dynamical masses of real groups. These are plotted in Fig. \ref{corr3d}. The distributions of all these estimates for each group are summarised as box-plots, where the median is shown as an orange vertical tick mark; the $16^{th}$ and $84^{th}$ percentiles by the box edges; and the extrema by the whiskers. The slope of the correlation between the dynamical and real mass in each panel matches the corresponding value of $\eta$ for each radius. We compute this slope by a least-squares linear regression fit in each case.
We see a clear correlation present at all radii, with a constant relative uncertainty of $\delta \eta/ \eta \approx0.12$, an indication that the correlation still holds even when including stars at the group edges.

\begin{figure*}
        \centering
        \includegraphics[width=0.99\textwidth]{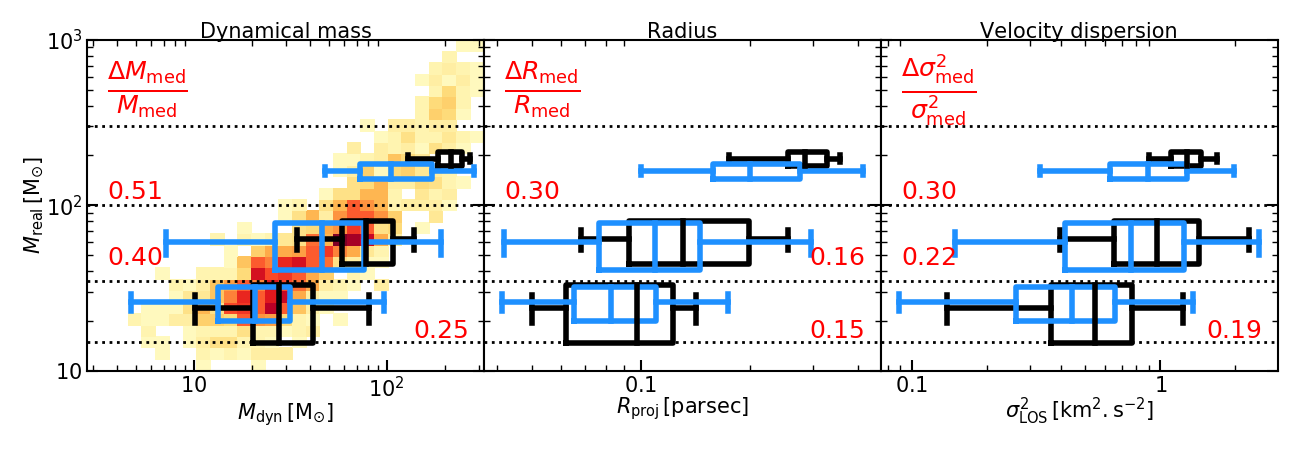}
    \caption{Comparison between projected and real group properties for three mass bins: $[15; 35]\, \rm M_\sun$,  $[35; 100]\, \rm M_\sun$ and $[100; 300]\, \rm M_\sun$ (dashed lines). Distributions of each estimate are represented as black box plots for real groups and blue ones for projected groups. The relative error on the median for each distribution is annotated in red. In each box plot, the vertical middle tick represents the median of the distribution, the edges show the $16^{th}$ and $84^{th}$ percentiles, and the whiskers correspond to the full extension of the distribution.
    The estimates plotted correspond to the $90\%$ Lagrangian radii but other radii give similar trends. The height of the boxes is proportional to the number of groups inside each mass bin. The 2D histogram of Fig. \ref{corr2d} is over-plotted on the panel of the dynamical mass estimate.}
    \label{underestim}
\end{figure*}

\begin{figure*}
        \centering
        \includegraphics[width=0.99\textwidth]{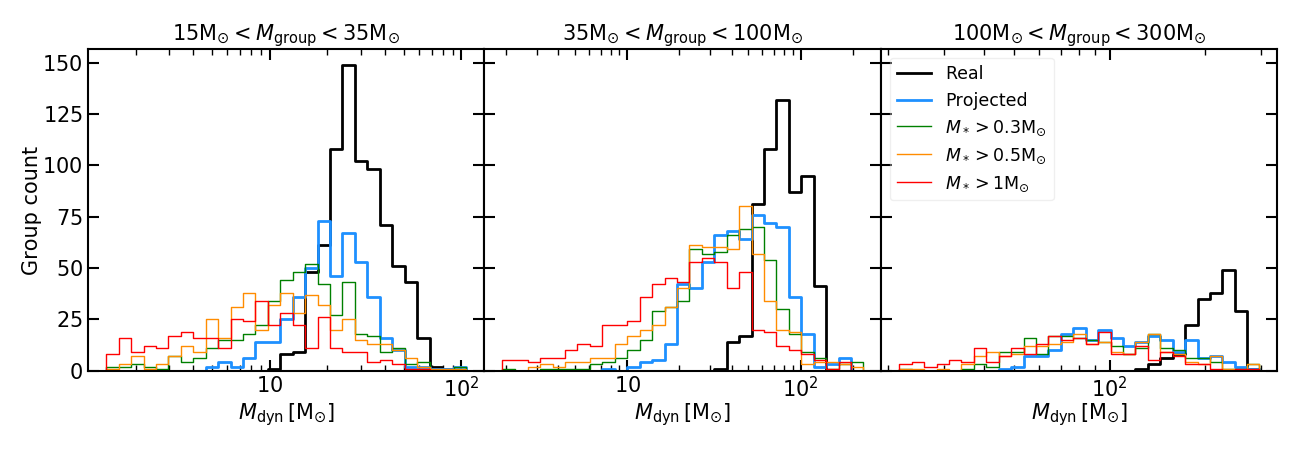}
    \caption{Comparison of the dynamical mass estimated on projected groups with different levels of incompleteness. The three panels correspond to the different real mass bins chosen in Fig. \ref{underestim}. The black histogram is the reference set of real groups and the blue histogram is the set of complete projected groups. The histograms in green, orange and red represents projected groups that are only complete down to $0.3\, \rm M_{\sun}$, $0.5\, \rm M_{\sun}$ and $1\, \rm M_{\sun}$ respectively. The figure shows the results using the $90\%$ Lagrangian radii.}
    \label{incomplete}
\end{figure*}

\subsection{Projected groups} \label{projG}

We focus now on the projected groups. We re-use the $96$ viewing angles of the whole region computed in \S\ref{Mdyn} to reach a similar number of mass estimates. As changing the line-of-sight will result in a different identification of projected groups, we re-compute the MST in each projection. Small variations in the number of groups identified in each projection lead to a different total number of projected groups of $1425$. Their dynamical masses were estimated with the corresponding $\eta$ factors constrained above (\S\ref{Mdyn}). The correlation with their real masses is plotted in Fig. \ref{corr2d} as a two-dimensional histogram. The black line shows the correlation derived from real groups. 
However, all mass estimates are shifted to lower values, falling to the left of the one-to-one relation on the figure. This highlights the fact that the mass of a group is underestimated when viewed in projection, irrespective of its true mass. The trend is not different from one panel to another, confirming that it is not sensitive to the choice of the radius.
Globally, the deficit is greater than $50\%$ for more than half of the groups, reaching $100\%$ for a quarter of them. We attribute this systematic error to the misidentification of clumps due to projection effects. As all radii are affected, the centre and the outskirts of projected groups are similarly impacted.

To  further investigate the reason for this deviation, we also compared radii and velocity dispersions of real and projected groups with similar masses. We divided both sets into three bins in terms of  mass: $[15; 35]\, \mathrm{M_\sun}$,  $[35; 100]\, \mathrm{M_\sun}$ and $[100; 300]\, \mathrm{M_\sun}$. The results corresponding to the $90\%$ Lagrangian radii are plotted in Fig. \ref{underestim}, but other radii give similar trends. We take a statistical approach to explore differences between real and projected groups. The distributions within each mass bin are represented by box plots of different colours: black for real groups and blue for projected groups. We recover on the left panel the systematic error in dynamical mass between real and projected groups (previously seen on Fig. \ref{corr2d}). The two other panels show the respective distributions of radii and velocity dispersions used to compute the dynamical mass. 
The properties of projected groups in a given mass bin are statistically lower than those of real groups. Moreover, the trend is more pronounced for the most massive groups. 
For the less massive groups, their Lagrangian radii are globally comparable whereas their velocity dispersions are systematically about $20\%$ lower. We note here that this is not the case when choosing only the stars lying in the $50\%$ or $75\%$ Lagrangian radii where Poissonian fluctuations can increase the difference up to $\approx30\%$.
For groups more massive than $100\MSun$, both radii and velocity dispersions are equally shifted by about $30\%$, independently of the chosen radius.

In summary, projection effects lead to significant underestimation of both the size and velocity dispersion of projected groups. Only the radii of the less massive groups are nearly not affected.
Some may find it surprising that the velocity dispersion is lowered in projected groups where some stars are not part of any clumps. However, we must bear in mind that these stars are in the same region, and therefore they are likely to have a velocity close to the mean value (of $\approx0\, \rm{km/s}$ in our case, see \S\ref{velocinfo}), which is the reason for the lower velocity dispersion in projected groups.


\subsection{Incompleteness}
Projection effects are always present in astronomical observations where spherical symmetry is a valid ansatz. This eases comparison with theoretical models. However, this assumption is not justified for young systems.
We indeed show that a lack of precise information in the line of sight can introduce systematic biases in the clump properties.
        
Another common bias when studying young stellar systems arises from incompleteness of the target populations. This also affects particularly star forming regions where extinction can be significant. We may estimate the importance of the completeness of the sample in comparison with that of the morphological projection biases that we have been discussing so far. At first order, the faintest sources are also the least massive ones, and so we apply different mass cuts to the set of projected groups (at $0.3\, \mathrm{M_\sun}$, $0.5\, \mathrm{M_\sun}$ and $1\, \mathrm{M_\sun}$). Again, we estimated the properties of the resulting groups on the different lines of sight. Figure \ref{incomplete} shows the resulting dynamical mass distribution within the three mass bins defined earlier in Fig. \ref{underestim}. Distributions of real groups and complete projected groups from Fig. \ref{underestim} are also plotted for comparison. We observe that the distribution of dynamical mass estimates is progressively shifted to smaller values when incompleteness increases. This trend is expected, because the more massive stars selected are segregated in space over a smaller volume \citep{dorval2016}, while their velocity dispersion will also be lower due to star--star interactions. 
As incompleteness of the sample will always increase the scatter owing to lower number statistics, dynamical mass distributions of incomplete groups show greater dispersion.

This analysis is just the first step in a longer process, as we know that the problem of incompleteness is complex. Detection limits of observations is wavelength dependant and extinction is never constant inside the same star forming regions. Indeed, dust clouds also harbour filamentary structures in interaction with newly born stars and dramatically modify the aspects of these clusters. Other important additions are the stellar photometry and interstellar reddening, two quantities affecting the sampling of stars with opposite results. Indeed, the most massive stars are the brightest but are also likely to be the most embedded \citep{motte2018}. A self-consistent simulation including hydrodynamics would be needed to more accurately explore these aspects.

\section{Discussion} \label{discussion}

\subsection{Decontamination} \label{velocinfo}

We recall that the stars contaminating the set of projected groups are stars from the same region, separated only by a few parsecs and sharing the same global dynamical state. Therefore, it can be difficult to distinguish between contaminants and group members based on their kinematics alone. We ask if \textit{Gaia} DR2 proper motions can separate two groups seen in projection using their individual proper motions.
If we use a reference distance to Orion of $400\, \rm{pc}$ \citep{kounkel2018}, the highest accuracy of \textit{Gaia} DR2 proper motions is $0.1\, \rm{km/s}$ at $M_{\rm G}=13$ but only reaches $1\, \rm{km/s}$ at $M_{\rm G}=18$. As \textit{Gaia} EDR3 will improve this value by a factor of two, we can expect to distinguish between two groups separated by a systematic motion of more than $0.5\, \rm{km/s}$.

\begin{figure}
    \centering
        \includegraphics[width=0.99\columnwidth]{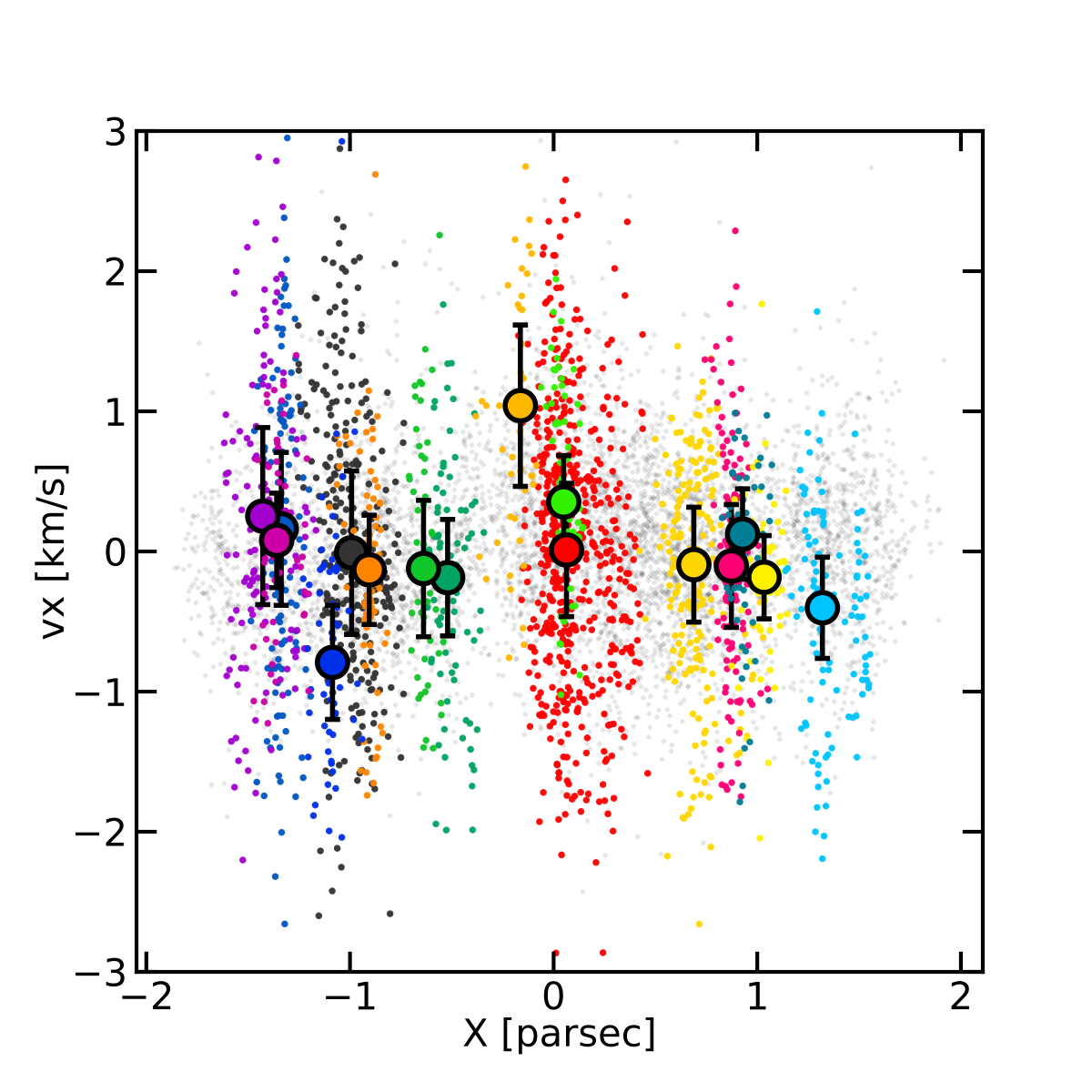}
    \caption{Position--velocity plot of stars in the sample. Stars that do not belong to any group are coloured in light grey whereas those in real 3D groups match the colours used in Fig. \ref{MST3D}. The filled circles are the mean values for each group and the error-bars represent the standard deviation.}
    \label{p-v_plot}
\end{figure}

Figure \ref{p-v_plot} shows a one-dimensional position--velocity plot of stars in real (3D) groups. Each point is colour-coded as in Fig.~\ref{MST3D}. Field stars are shown in light grey.  
The filled circles are the mean values for each group and the error-bars represent the standard deviation, which is an estimate of the typical internal motion of the  clumps. An indication of the relative motions between these groups is given by the standard deviation of these mean values, which is $\approx0.4 \rm{km/s}$. This is of the same order as the internal motion, which is $\approx 0.5\,\rm{km/s}$. 
We conclude that the majority of real groups cannot be differentiated. We find only one instance where two out of sixteen groups are separated by $\approx 2 \rm{km/s}$ which is within the \textit{Gaia} detection limit.
This is by construction a feature brought by our GDF method, because we did not introduce large-scale motion in our fragmentation method. Such large-scale modes can be included in a future study. 

\subsection{Varying the total number of stars in the region} \label{lowN}
\begin{figure}
    \centering
        \includegraphics[width=0.8\columnwidth]{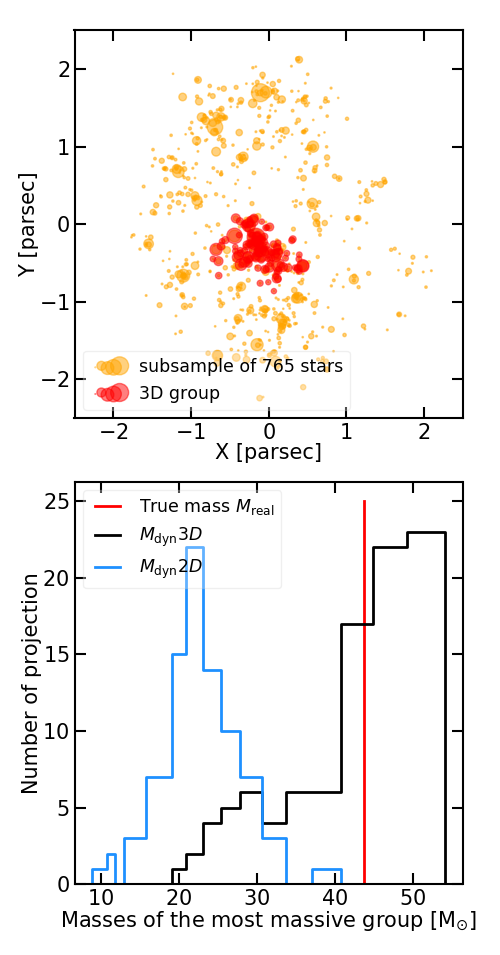}
    \caption{Upper panel: Sub-sample extracted from a GDF model of $1000$ stars with the method of \S\ref{HOP}. Only one major group is identified in 3D by the MST (red dots).
    Lower panel: Distributions of the dynamical mass estimated on different projections of the main group identified by the MST in 3D (black line) and in projection (blue line). The true mass is marked as the vertical red line.}
    \label{GDF1000}
\end{figure}

Although the reference set of $15000$ stars we previously studied allows us to explore clumps of different membership and density, the large projected density of $\approx500\, \rm{stars/pc}^2$ can have an impact on the level of contamination of these clumps. Moreover, such clumpy configurations are not always seen in star forming regions. However, it is clear that none of the star forming regions are perfectly isolated \citep{kuhn2014}. The local environment of a particular region, even if it is not as structured as in our reference model, can still introduce some biases when inferring its properties. 
To probe such conditions, we built a GDF model with only $1000$ stars and repeated the analysis. The sub-selection method with the \texttt{HOP} algorithm (\S\ref{HOP}) extracts a sample of $765$ stars for a surface density of $50\, \rm{stars/pc}^2$. This new model is, as expected, less structured: only one major over-density developed at the centre of the region (see upper panel of Fig.~\ref{GDF1000}). We therefore focus on this single clump detected by the MST. However,  some structures exist in its environment that may contaminate it when seen in projection. We performed the estimation of the dynamical mass of this single group as in \S\ref{Mdyn}. We used the values of $\eta$ constrained previously in \S\ref{Mdyn}. We also identify groups in projection for the same $96$ lines of sight\footnote{We note here that the MST identifies sometimes more than one group in projection. In such cases, we only retain the most massive one.} and estimate their dynamical mass too. Finally we compare the distribution of both estimations. The distributions corresponding to the $75\%$ Lagrangian radii can be seen in the bottom panel of Fig. \ref{GDF1000}. Other radii show similar behaviours. The true mass of the group is marked as the red line for reference. 
The results are similar to those shown in Fig. \ref{incomplete}: the dynamical mass of the projected group is systematically lower than that of the real group. Therefore, even the smaller number of sub-structures in this model can still lead to inaccurate estimation of clump properties. 
Second, although the value of $\eta$ was not calibrated on this model, the dynamical mass is well retrieved in 3D.
Our overall conclusions are therefore not sensitive to the precise surface density at least from $50\, \rm{stars/pc}^2$ to $500\, \rm{stars/pc}^2$.

\subsection{Clump-finding algorithm}

Our study specifically aims to assess the bias introduced by projection effects within a star forming region. We show that project effects systematically lower the spatial and kinematic properties of clumps. However, the amplitude of this shift is difficult to quantify. The variation that we estimate to range from $10$ to $30\%$  is due to the intrinsic nature of projection bias, which depends on the  shape and orientation of a particular region. It is also linked to the method we choose. 

However, most clump-finding algorithms also use projected spatial coordinates.
For instance \cite{joncour2018} proposed a method to calibrate the DBSCAN algorithm using a statistical approach. The DBSCAN is a close alternative to the MST. These latter authors exploit the nearest-neighbour statistics of a 2D Poissonian distribution to compute the parameters of the DBSCAN. This allows them to select only projected over-densities at three-sigma level over random expectation.
Nevertheless, this method is most suitable for small-scale structures of approximately ten members, where contamination in the line of sight is weak.
For bigger clumps, like those that we identified in the present study, projection effects are more significant. Therefore, we may expect that such a method will lead to similar results to the MST approach adopted here.

A step forward would be to build a set of GDF models reproducing configurations of a wide range in density, and train a machine-learning algorithm on this sample to bridge the 3D identifications to 2D projections with a significance estimator, similarly to what was developed by \cite{joncour2018} for DBSCAN. We show here that all groups show comparable projection biases, with a trend of increasing amplitude with group mass. This suggests that a data set of models with reasonably small groups ---of up to 1000  stars--- should still yield useful statistics while remaining manageable in terms of volume. 
Further work in that direction is underway.

\subsection{The mass-to-light ratio $\eta$} \label{eta}

A secondary result emerging from this study is the value of the mass-to-light ratio  $\eta$ linking the true and dynamical masses of real groups (see \S\ref{Mdyn}). 
The value  $\eta \sim 10$ is often chosen to match a
wide range of spherical King-Mitchie models \citep{spitzer1987, mengel2002}. However, \cite{fleck2006} already showed that $\eta$ varies tremendously through the lifetime of a cluster, mostly as a result of mass segregation. 

\begin{figure}
    \centering
        \includegraphics[width=0.9\columnwidth]{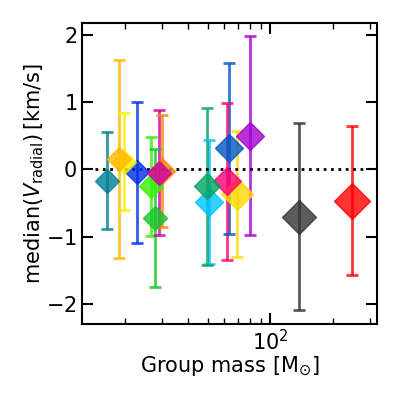}
    \caption{Median of the radial velocity of stars in real groups with respect to their centre of mass. The colours are chosen to match those in Fig. \ref{MST3D}.}
    \label{Vrad}
\end{figure}

The value $\eta \simeq 10$ is about a factor three greater than our evaluation of $\approx 3.4$ for real groups. Previous works (see \eg \citealt{spz2010, parker2016}) predicted that low values of $\eta$ could be a signature property of young systems, one related to their state of equilibrium.
As nothing suggests that real groups are in virial equilibrium, such a low value of $\eta$ could be an indication of expansion. 
We rule out this explanation by computing the radial velocities of stars in real groups from their respective centres of mass. The median for each group is plotted in Fig. \ref{Vrad}. It shows that real groups are in a quasi-steady state, with $\overline{v}_{\mathrm{r}} \simeq 0 \, \mathrm{km/s}$. This is perhaps surprising, because they are in close interaction with their environment, for instance through the tidal field of neighbouring groups and fly-by stars.
However, it must be borne in mind that these configurations are a snapshot taken after only $t \simeq 0.5\, \mathrm{Myr}$. The age of the system is of the same order as the dynamical crossing time for groups, when little internal kinetic energy diffusion has had time to drive expansion. 

Precise determination of $\eta$ should help observers to put further observational constraints on the dynamical state of sub-clusters in star forming regions. 
For instance \cite{kuhn2019} recently computed the \textit{virial-expected} velocity dispersion of $19$ young clusters. 
A lack of constraints on the value of $\eta$ has prevented confirmation of the equilibrium dynamical state of these systems. Assuming a value of $10$ lead \cite{kuhn2019} to conclude that all of these clusters were unbound, although their \textit{Gaia} proper motions did not show clear signs of expansion. Using a reduced value of $5$, closer to our evaluation, would reconcile these contradictory results. We stress that the systems that we have analysed, which are of linear scale $\simeq 0.5 \,\mathrm{pc}$ or less, are small compared to those of \cite{kuhn2019}. To make a more meaningful comparison with their survey would require
that we develop models of a full-size $\simeq 100 \,\mathrm{pc}$ region, with disc-shearing and the vertical disc structure of the Milky Way added to the GDF modelling of \S2. 
We hope to report on these aspects in a future contribution.

\section{Conclusions} \label{conclusion}

In the present study, we use GDF models  to self-consistently reproduce a fragmented distribution of stars to assess the systematic bias introduced by an imprecise identification of stellar clumps in projection compared to an ideal one in 3D. To this end, we performed a statistical comparison between the properties of stellar over-densities (of membership $N_{\rm stars}>50$) identified in 3D and in projection by a generic MST algorithm within a single star forming region of $\approx 4\, \rm{pc}$ in width\footnote{We recall that we did not take into account any additional foreground or background contamination besides the one introduced by the region itself.}. Our main results can be summarised as follows:
\begin{enumerate}
\item We highlight that  the dynamical mass of  clumps identified in projection is  systematically underestimated compared to the ideal 3D identification. This systematic error is greater than $50\%$ in about half of the cases and can reach up to $100\%$ in a quarter of them. This is due to the incorrect selection of the clump-finding algorithm when dealing only with projected coordinates in such fragmented regions.

\item This projection bias impacts both radius and velocity dispersion estimates of projected groups. Eliminating potential outliers at the edges of the groups identified by the MST did not lead to any improvements.

\item We notice a trend ranging from the properties of less massive ($M<100 \rm M_{\odot}$) groups, which are less  affected or even not at all, to those of the most massive groups ($M>100 \rm M_{\odot}$), which are more systematically affected. For these latter, radii and velocity dispersions are statistically $30\%$ lower. 

        
\item As the projection effect is not the only bias in star forming regions, we also considered simple prescriptions to account for incompleteness in the observations. We show that sample incompleteness increases the systematic error. However, its impact is of a different nature and can be seen as a blurring effect. For example, the most massive groups are almost unaffected since they still contain a sufficient number of massive members for them to be characterised.
        
\item We conclude that projection effects are the main limitation in estimating properties of clumps identified within a single fragmented region with no or insufficiently accurate distance estimates. For most nearby star forming regions, the best Gaia parallaxes have uncertainties of a few percent. For these  distant regions, that is, at more than 100 pc, this prevents any clump-finding algorithm from properly defining clumps of less than a few parsecs in size. Accurate identification of such clumps is a key challenge in better constraining the processes of star formation. This also has major implications for predicting the future of these young systems in bound clusters or unbound associations. 

\item We also show that GDF models offer an efficient method to initialise \nbody simulations of young clusters. With this method, we are able to explore the structures built by gravity alone in more detail in order to better understand the role of gravity compared to the many processes at play in star forming regions.

\end{enumerate}

\begin{acknowledgements}
We thank the referee for the constructive remarks which led to improve the quality of the paper. Timothé Roland also acknowledges Paolo Bianchini, Caroline Bot, Joe Lewis and Jonathan Chardin for their helpful comments and suggestions. This work has used the \texttt{AMUSE} framework\footnote{\url{https://github.com/amusecode/amuse}} \citep{portegies2018, amuse2013a, amuse2013b, amuse2009} and the \texttt{Python} programming language with the following packages: \texttt{matplotlib} \citep{hunter2007}, \texttt{NumPy} \citep{numpy2020}, \texttt{SciPy} \citep{scipy2020} and \texttt{Astropy} \citep{astropy2013, astropy2018}.
\end{acknowledgements}

\bibliographystyle{aa} 
\bibliography{biblio}

\end{document}